\documentclass[aps,prb,reprint,amsmath,amssymb,lettersize,showpacs]{revtex4-1}
\usepackage{graphicx}
\usepackage{dcolumn}
\usepackage{bm}
\usepackage{braket}

\begin{document}


\title{First-principles description of van der Waals-bonded spin-polarized systems using vdW-DF$+U$ method---application to solid oxygen at low pressure}



\author{Shusuke Kasamatsu}
\email{kasamatsu@issp.u-tokyo.ac.jp}
\author{Takeo Kato}
\author{Osamu Sugino}%
\affiliation{%
The Institute for Solid State Physics, the University of Tokyo \\
5--1--5 Kashiwanoha, Kashiwa-shi, Chiba 277-8571, Japan 
}%

\date{\today}
\begin{abstract}
The description of the molecular solid phase of O$_2$, especially its ground-state antiferromagnetic insulating phase, is known to be quite unsatisfactory within the local and semilocal approximations conventionally used in the Kohn-Sham formalism of density functional theory (DFT). The recently-developed van der Waals (vdW) density functionals, vdW-DF, that take into account nonlocal correlations have also shown subpar performance in this regard. The difficulty lies in the subtle balance between the vdW interactions and the exchange coupling between the spin-triplet state of molecules in the molecular crystal. Here, we report that the DFT$+U$ approach used in combination with the vdW-DF performs surprisingly well in this regard, and discuss the reasoning behind this behavior. We also apply this approach to study the recently-reported magnetic field-induced $\theta$ phase of solid O$_2$.
\end{abstract}


\maketitle


\section{Introduction}
Solid oxygen is unique in that it is a molecular crystal comprised of spin-polarized molecules. Because the van der Waals (vdW) interaction and the magnetic interaction between the O$_2$ molecules are comparable in magnitude and compete with each other, solid O$_2$ exhibits strong spin-lattice coupling. This leads to a variety of structural/magnetic phases under varying pressures and temperatures \cite{Uyeda1985,Freiman2004}.  In addition, recent advances in high-power magnets have opened up the possibility of exploring phase transitions induced by magnetic fields, and indeed, Nomura and coworkers have reported a new phase of solid oxygen at a magnetic field of $\gtrapprox 100$ T \cite{Nomura2014,Nomura2015}. 
Due to the difficulty in experimental setup, it is currently impossible to determine the structure and various physical properties at such high magnetic fields, and first-principles simulations are expected to help in this regard.

\begin{figure}
\centering
\includegraphics[width=0.9\columnwidth]{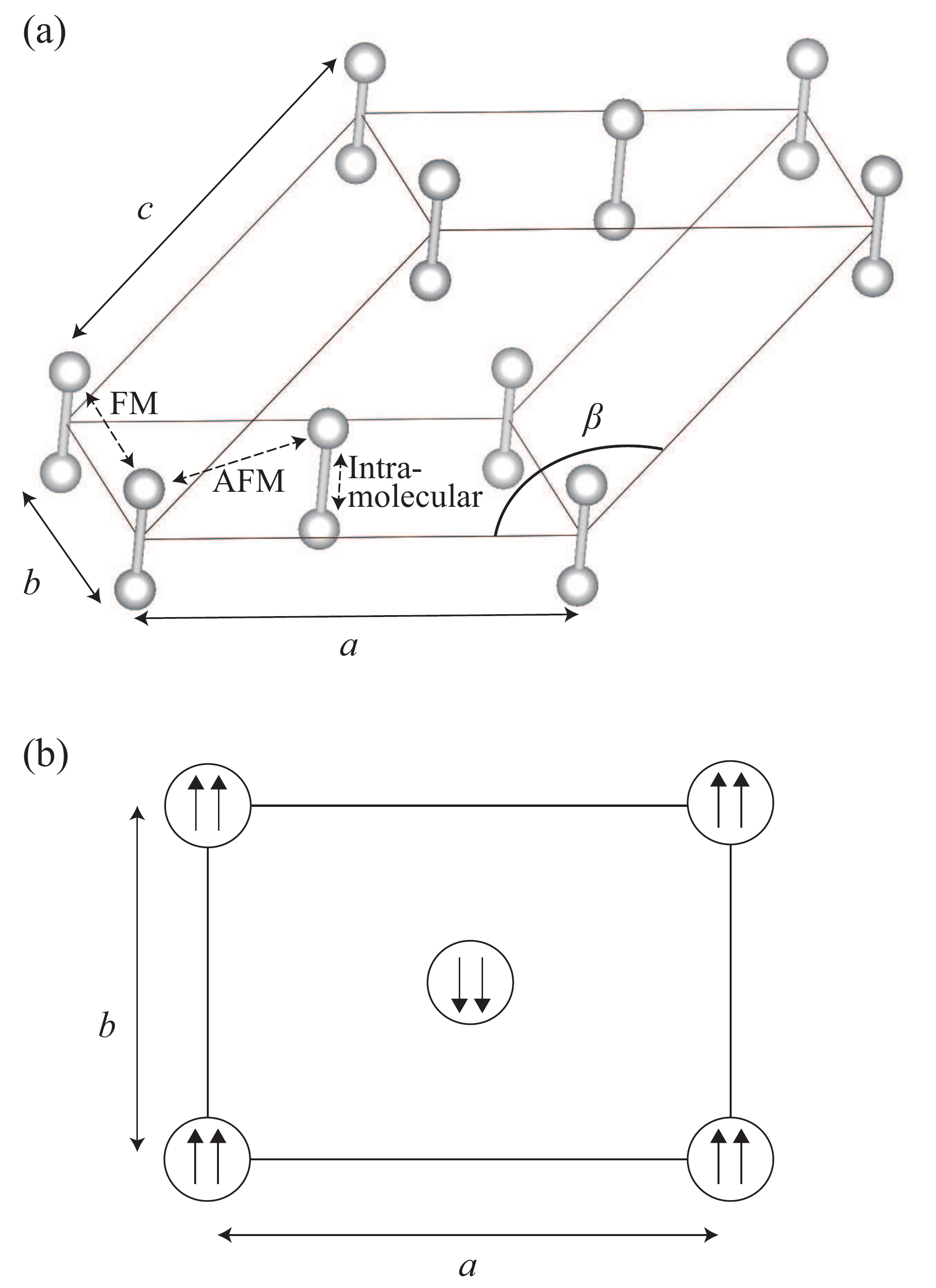}
\caption{\label{fig:alphaO2} (a) The structure of the low-temperature monoclinic $\alpha$ phase of solid oxygen with lattice parameters $a$, $b$, $c$, and the monoclinic angle $\beta$. The intramolecular and intermolecular (anti)ferromagnetic bonds investigated using pCOHP analysis in Sec.~III are indicated by dashed arrows. (b) Schematic of the antiferromagnetic order in the basal $ab$ plane.}
\end{figure}

However, due to the subtle balance of vdW and exchange interactions mentioned earlier, satisfactory description of solid O$_2$ from first principles is quite challenging; in fact, it may be considered one of the most critical benchmarks for measuring the predictive capability of electronic structure methods \cite{Obata2015a}. The local and semilocal approximations to the density functional [e.g., the local spin density approximation (LSDA) and the generalized gradient approximation (GGA)] used conventionally in the Kohn-Sham method of density functional theory (DFT) fail rather miserably in predicting the structure of the low-temperature ground state monoclinic ($\alpha$) phase (Fig.~\ref{fig:alphaO2}). This is not surprising because of the nonlocal nature of the vdW interaction, but even sophisticated functionals with nonlocal correlations show small improvements. For example, the vdW-DF functional of Langreth, Lundqvist, and coworkers \cite{Dion2004} has been shown to perform adequately in terms of predicting the volume of the unit cell, but the problem remains in predicting the shape of the unit cell; the calculated lattice parameters are off by as large as 20\% compared to experiment \cite{Obata2015}. 
The poor performance of these functionals was suggested to be due to overestimation of bonding in the antiferromagnetic molecule pairs compared to ferromagnetic pairs \cite{Obata2015,Obata2015a}. The magnetic interaction $J$ is proportional to $-t^2/\Delta E$, where $t$ is the transfer integral between sites, $\Delta E$ is the energy gap separating the spin-up and spin-down states sandwiching the Fermi level, and $J$ is taken to be negative for antiferromagnetic coupling. GGAs are known to underestimate $\Delta E$ and overestimate $|t|$, so it is not surprising that the antiferromagnetic interaction is overestimated. vdW-DF does not improve much in this regard, so Obata and coworkers opted to consider a spin-polarization dependent gradient correction to be used in combination with vdW-DF \cite{Obata2015, Obata2015a}. In their approach, two scaling parameters for relative spin polarization and spin-dependent gradient correction were introduced in the vdW-DF-SGC method, and the errors in the magnetic interaction were corrected to some extent depending on the chosen parameters. However, optimizing the
two parameters still did not yield completely satisfactory results for the lattice parameters of the $\alpha$ phase.

Aside from DFT simulations, prediction of the structure using intermolecular forces parametrized from configuration interaction calculations of the O$_2$--O$_2$ molecule dimers has shown some success \cite{Nozawa2002,Nozawa2008}. Lattice dynamics simulations have also succeeded in a rather good description of the $\alpha$ and $\delta$ phases \cite{Etters1985}. However, the transferability of such description of solid O$_2$ to other phases is questionable, especially considering the richness of the oxygen phase diagram encompassing antiferromagnetic, ferromagnetic, and paramagnetic insulating states as well as metallic states.

In this work, we consider the simpler approach than, e.g., vdW-DF-SGC of making use of the DFT$+U$ approach, which takes into account strong on-site interactions that aren't treated properly in LSDA and GGA by adding a
Hubbard-$U$ like term to the energy functional (see e.g., Ref.~\onlinecite{Kulik2015} for a recent review). The approach has seen much use on $3d$ and $4f$ states in transition metal oxides or molecular complexes, and recent studies have shown its effectiveness also on the oxygen $2p$
states in oxides \cite{Park2010,Plata2012}. Here, we consider its application to the molecular crystal of `pure' oxygen in combination with vdW-DF. Since the DFT$+U$ approach is known to increase the energy band gap and favors localization of electrons, it
may be considered a natural choice for correction of the error in $J \propto -t^2/\Delta E$ mentioned above.

There are several flavors of DFT$+U$ implementations; here, we employ the simplified rotationally invariant version by Dudarev {\it et al.} \cite{Dudarev1998}, which employs only one parameter, $U_\text{eff}$, in the
description of on-site repulsion. That is,
\begin{equation}
E_{U} = \frac{U_\text{eff}}{2} \sum_\sigma \left [ \left( \sum_{m_1} \hat{n}^\sigma_{m_1,m_1} \right ) - \left (\sum_{m_1,m_2} \hat{n}^\sigma_{m_1,m_2} \hat{n}^\sigma_{m_2,m_1} \right ) \right ]
\end{equation}
is added to the total energy functional, where $\hat{n}$ is the on-site occupancy matrix of oxygen $p$ states.
The vdW-DF exchange correlation energy is written as 
\begin{equation}
E_\text{xc}^\text{vdW-DF} =  E_\text{x}^\text{GGA} + E_\text{c}^\text{LDA} + E_\text{c}^\text{nl},
\end{equation}
where the first term is GGA exchange, the second term is  LDA correlation, and the last term is the nonlocal correction based on the plasmon picture \cite{Dion2004}. 
We test the original vdW-DF which employs revPBE \cite{Zhang1998} for $E_\text{x}^\text{GGA}$ \cite{Dion2004}. We also test the vdW-DF with optB86b exchange \cite{Klimes2011}, which has been shown to produce results that are in general more accurate than the vdW-DF with revPBE exchange \cite{Berland2015}. 
We note that in the current work, nonlocal correlation $E_\text{c}^\text{nl}$ does not depend on the spin density and is evaluated from the sum of the spin-up and spin-down densities. The influence of the spin density enters explicitly only through the exchange functional $E_\text{x}$. Strictly speaking, this implementation cannot be justified based on the original vdW-DF derivation as it ignores the fact that spin changes the plasmon dispersion, but it has still been used pragmatically (e.g., Ref.~\onlinecite{Vanin2010}). On the other hand, a fully consistent vdW functional including spin-dependence of the nonlocal correlation (svdW-DF) has recently been proposed \cite{Thonhauser2015}, and it may indeed play a role in describing this system. However, as we show below, the main culprit in the unsatisfactory description of this system is the well-known electron delocalization error in semilocal density functional approximations. svdW-DF does not correct for this, at least not explicitly, so we tentatively suggest that svdW-DF will show minor improvement in this system.

The choice of the value of $U_\text{eff}$ also deserves attention. One may consider it a correction for the lack of derivative discontinuity in semilocal density functionals and determine its value using either a linear response \cite{Cococcioni2005} or a self-consistent procedure \cite{Kulik2006}. On the other hand, much of the literature on DFT$+U$ takes $U_\text{eff}$ to be a tuning parameter for reproducing certain properties such as the band gap or cohesive energies. We take the latter approach in this work, focusing on the structure of solid oxygen and the physics of how the $U_\text{eff}$ parameter affects this system.
We apply this vdW-DF$+U$ approach to calculate the lattice parameters of the $\alpha$ phase ($\alpha$-O$_2$). We find that this vdW-DF$+U$ approach yields surprisingly good results in reproducing the experimental lattice parameters of $\alpha$-O$_2$ when the single parameter $U_\text{eff}$ is optimized. 
To understand this effect, we examine the effect of the Hubbard term on the electronic structure and the inter/intramolecular bonding of oxygen.
Finally, we apply this method to examine the candidate cubic Pa$\bar{3}$ structure for the magnetic-field induced $\theta$ phase and discuss whether this is indeed justifiable as the realized structure at magnetic field $B > 100$ T.

\section{Methodology}
The calculations are performed using VASP \cite{Kresse1996,Kresse1996a} code based on the Kohn-Sham formalism of density functional theory (KS-DFT) \cite{Hohenberg1964,Kohn1965}. The projector-augmented wave (PAW) method \cite{Blochl1994} is used to describe ion-electron interactions, and the wave functions are expanded by a plane wave basis set with a cutoff energy of 2000 eV. The structural relaxations are performed until forces on each ion become smaller than $10^{-3}$ eV/\r{A}. 
The LSDA, GGA-PBE, and vdW-DF approximations to the density functional, as well as the combination of vdW-DF and the $+U$ approach are tested on solid O$_2$. A $5 \times 6 \times 5$ $\mathbf{k}$-point mesh is employed for the single unit cell of the monoclinic $\alpha$ phase, while $9 \times 9 \times 9$ $\mathbf{k}$-point mesh is employed for the single unit cell of the cubic $\theta$ phase and a $3 \times 3 \times 3$ mesh is employed for phonon calculations in the $2 \times 2 \times 2$ expanded supercell of the $\theta$ phase. We also employed a monoclinic unit cell expanded by a factor of two in the $b$ direction with a slightly denser mesh of $8 \times 5 \times 8$ for the bonding analysis described below. The finite displacement method \cite{Kresse1995, Parlinski1997} was used for phonon calculations using phonopy package \cite{phonopy} for pre- and post-processing of VASP input and output files.

To quantify the bonding strength between O$_2$ molecules, we employ the LOBSTER-2.1.0 code \cite{Maintz2016} for performing projected crystal orbital Hamilton population (pCOHP) analysis. The crystal orbital Hamilton population (COHP) is defined as \cite{Dronskowski1993}
\begin{equation}
\text{COHP}_{\alpha,\beta} (\epsilon) = H_{\alpha,\beta} \sum_j u^\ast_{\alpha,j} u_{\beta,j} \delta (\epsilon_j-\epsilon),
\end{equation}
where $\alpha$ and $\beta$ refer to site-localized orbitals, $j$
specifies the band index, $H_{\alpha,\beta}$ is the Hamiltonian matrix
element, $u_{\alpha,j}$ and $u_{\beta,j}$ are the wave function
coefficients, and $\epsilon$ is the Kohn-Sham eigenenergy. A negative
(positive) $\text{COHP}_{\alpha,\beta}(\epsilon)$ value corresponds to a
bonding (anti-bonding) interaction. By summing up the COHP over all
$\alpha$ and $\beta$ belonging to an atom pair and  integrating up to
the Fermi level ($\epsilon_\text{F}$), one obtains the integrated COHP ($\text{ICOHP}(\epsilon_\text{F})$), which
corresponds roughly to the idea of bond order in molecular orbital theory or to
the transfer integral in the Hubbard-based models. To apply COHP analysis to the results of plane-wave DFT codes, one first needs to project the Kohn-Sham wave functions onto localized auxiliary basis functions, then perform similar calculations to obtain the pCOHP, the projected variant of COHP. It should be noted that the quality of the pCOHP depends on the quality of the projection (i.e., how well the projected wave functions reproduce the original wave functions), which can be evaluated by absolute charge spilling defined in Ref.~\onlinecite{Maintz2016}. All projection results presented in this paper have absolute charge spilling of less than 1.1\%.
In passing, it should be noted that (p)COHP accounts only for covalent-like bonds in the region where orbital wave functions overlap with each other; it does not account for, e.g., ion-ion Coulomb interactions \cite{Dronskowski1993}. We may expect short-range parts of the vdW interactions to be included in the pCOHP, but not the long-range interaction between parts without orbital overlap.

\begingroup
\begin{table*}
  \centering
  \caption{The volume $V$, lattice constants $a$, $b$, $c$, and $\beta$, the intramolecular bond length $l_{\text{O}_2}$, and the bulk modulus $B$ of $\alpha$-O$_2$ calculated using various functionals compared to experiment.}
\begin{ruledtabular}
    \begin{tabular}{lccccccc}
            & $V$ (\r{A}$^3$) & $a$ (\r{A}) & $b$ (\r{A}) & $c$ (\r{A}) & $l_{\text{O}_2}$ (\r{A}) &  $\beta$($^\circ$) & $B$ (GPa) \\
\hline
    Experiment \cite{Meier1984} & 69.5    & 5.4     & 3.43    & 5.09  & 1.28 & 133    & $\sim 6$ \cite{Akahama2001} \\
    LSDA & 42.0    & 3.92    & 2.95    & 4.15  & 1.20   & 119    & -- \\
    GGA-PBE & 75.4    & 4.21    & 4.18    & 4.90   & 1.22  & 119     & 1.2 \\
    vdW-DF-revPBE & 65.9    & 4.54    & 3.80    & 4.44  & 1.23  & 121     & 5.0 \\
    vdW-DF-revPBE \cite{Obata2015a} & 66.1    & 4.68    & 3.68    & 4.7  & 1.23  & 125     & --  \\
    vdW-DF-revPBE$+U$ ($U_\text{eff}=5$ eV) & 74.1    & 5.35    & 3.6     & 5.01  & 1.25 & 130     & 4.7 \\
    vdW-DF-optB86b & 48.6    & 3.59    & 3.58     & 4.19  & 1.22   & 115    & -- \\
    vdW-DF-optB86b$+U$ ($U_\text{eff}=12$ eV) & 69.7    & 5.29    & 3.48     & 5.01 & 1.27   & 131     & 4.4 \\
    vdW-DF-SGC \cite{Obata2015} & 75.7    & 5.43    & 3.61    & 4.57   & -- & 122 &  -- \\

    \end{tabular}%
\end{ruledtabular}
  \label{Tab:alpha}%
\end{table*}%
\endgroup

\section{Results and discussion}

\subsection{Effect of the $U_\text{eff}$ parameter on the structure of the $\alpha$ phase}

Table \ref{Tab:alpha} shows the lattice parameters of the $\alpha$ phase obtained by performing a variable-unit cell optimization procedure starting from the 
experimental structure using LSDA, GGA-PBE, vdW-DF, and vdW-DF-SGC functionals, as well as the vdW-DF$+U$ 
approach with an optimized $U_\text{eff}$ parameter. The bulk moduli are obtained by fitting energy vs. volume 
curves to Birch-Murnaghan equation of state. LSDA gives disastrous results on all fronts; the $\alpha$ structure is not 
even locally stable, the volume is grossly underestimated, and a nonmagnetic ground state is predicted. GGA-PBE 
and vdW-DF-revPBE give comparable results, with the former overestimating the volume and the latter 
underestimating the volume slightly. Even though the GGA-PBE functional contains no truly nonlocal correlation, error 
cancellation seems to result in equilibrium volume comparable to the vdW-DF-revPBE functional. There is one caveat, however,
as GGA-PBE
underestimates the bulk modulus, i.e., predicts a much softer lattice than experiment, while the vdW-DF-revPBE functional
predicts a bulk modulus that is much closer to experiment. The vdW-DF-optB86b functional seems 
to perform worse than vdW-DF-revPBE in that the underestimation of the volume is much more severe.
This is in line with the general trend in vdW-bonded systems that vdW-DF-revPBE predicts larger lattice constants than vdW-DF-optB86b, although usually, the optB86b predicts bonding distances closer to experiment \cite{Berland2015}.
These functionals correctly predict an antiferromagnetic ground state, but they all fail to reproduce the $b/a$ ratio, which is a measure of the exchange interaction in the $ab$ plane. There is also a noticeable underestimation of the $c$ parameter and the monoclinic angle $\beta$, which presumably originates from insufficient description of the exchange interaction between $ab$ planes. As mentioned in Sec.~I, the vdW-DF-SGC functional with two adjustable parameters show some improvement over vdW-DF-revPBE. Surprisingly, the vdW-DF$+U$ approach with only one adjustable parameter $U_\text{eff}$ shows even further improvement, reproducing the lattice constants within 2.1\% of experimentally reported values when the $U_\text{eff}$ value is optimized for the vdW-DF-optB86b and within 4\% for the vdW-DF-revPBE. 

\begin{figure}
\centering
\includegraphics[width=\columnwidth]{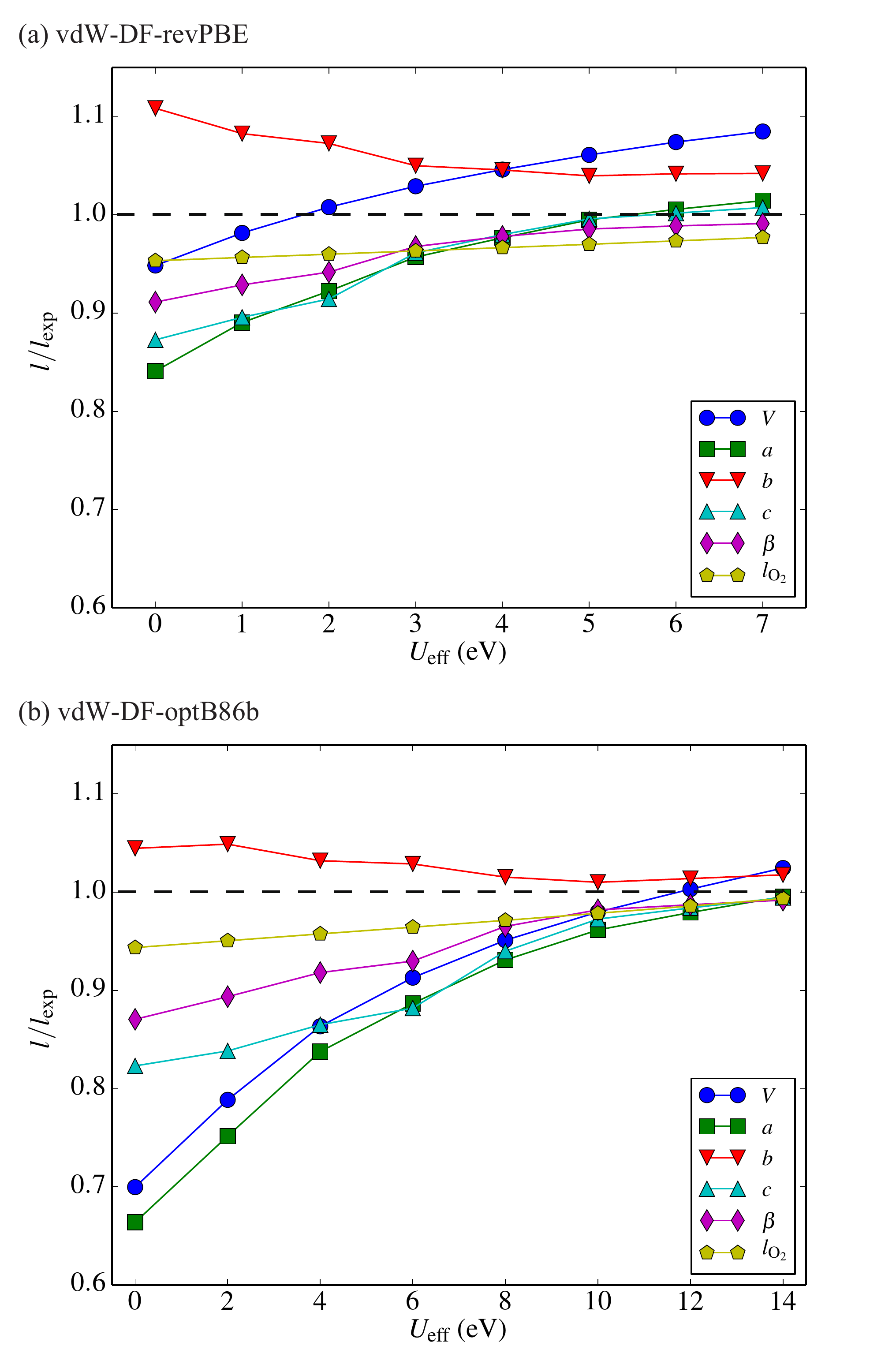}
\caption{\label{fig:U-lat} The ratio of the calculated lattice constants of $\alpha$-oxygen ($l = V, a, b, c, \beta, l_{\text{O}_2}$) vs.~the corresponding experimental values ($l_\text{exp}$) plotted as a function of the $U_\text{eff}$ for vdW-DF-revPBE (a) and vdW-DF-optB86b (b) functionals.}
\end{figure}

The effect of the $U_\text{eff}$ parameter on the calculated lattice parameters of the antiferromagnetic $\alpha$ phase is shown in Fig.~\ref{fig:U-lat}. At $U_\text{eff}=0$ eV, both vdW-DF-revPBE and vdW-DF-optB86b underestimate the volume, 
the latter more so. The $b$ parameter is overestimated while the $c$ and
$a$ parameters are underestimated in both functionals, most likely due
to the error in the exchange interactions mentioned above. We also note
that the monoclinic angle $\beta$ is underestimated. In addition, the
internal degree of freedom, i.e., the intramolecular O--O distance
$l_{\text{O}_2}$ is also underestimated. As the $U_\text{eff}$ parameter is increased from zero, all of the above-mentioned errors decrease. In the vdW-DF-revPBE functional, the error in the calculated volume becomes larger at above $U_\text{eff}=2$ eV while the errors in the other parameters continue to decrease up to $U_\text{eff} \sim
5$ eV. On the other hand, in the vdW-DF-optB86b functional, the errors in all lattice parameters continue decreasing up to $U_\text{eff} \sim 12$ eV. Although the vdW-DF-optB86b seems to perform worse compared to vdW-DF-revPBE at $U_\text{eff} = 0$ eV, it gives a much better result when the $U_\text{eff}$ parameter is optimized, in line with the general trend that optB86b exchange gives better results than revPBE. The rather high $U_\text{eff}$ value compared to most of the literature of transition metal systems may be due to lack of screening by conduction electrons in this system.

\subsection{Effect of the $U_\text{eff}$ parameter on the electronic structure and O$_2$--O$_2$ interaction}

\begin{figure*}
\centering
\includegraphics[width=1.75\columnwidth]{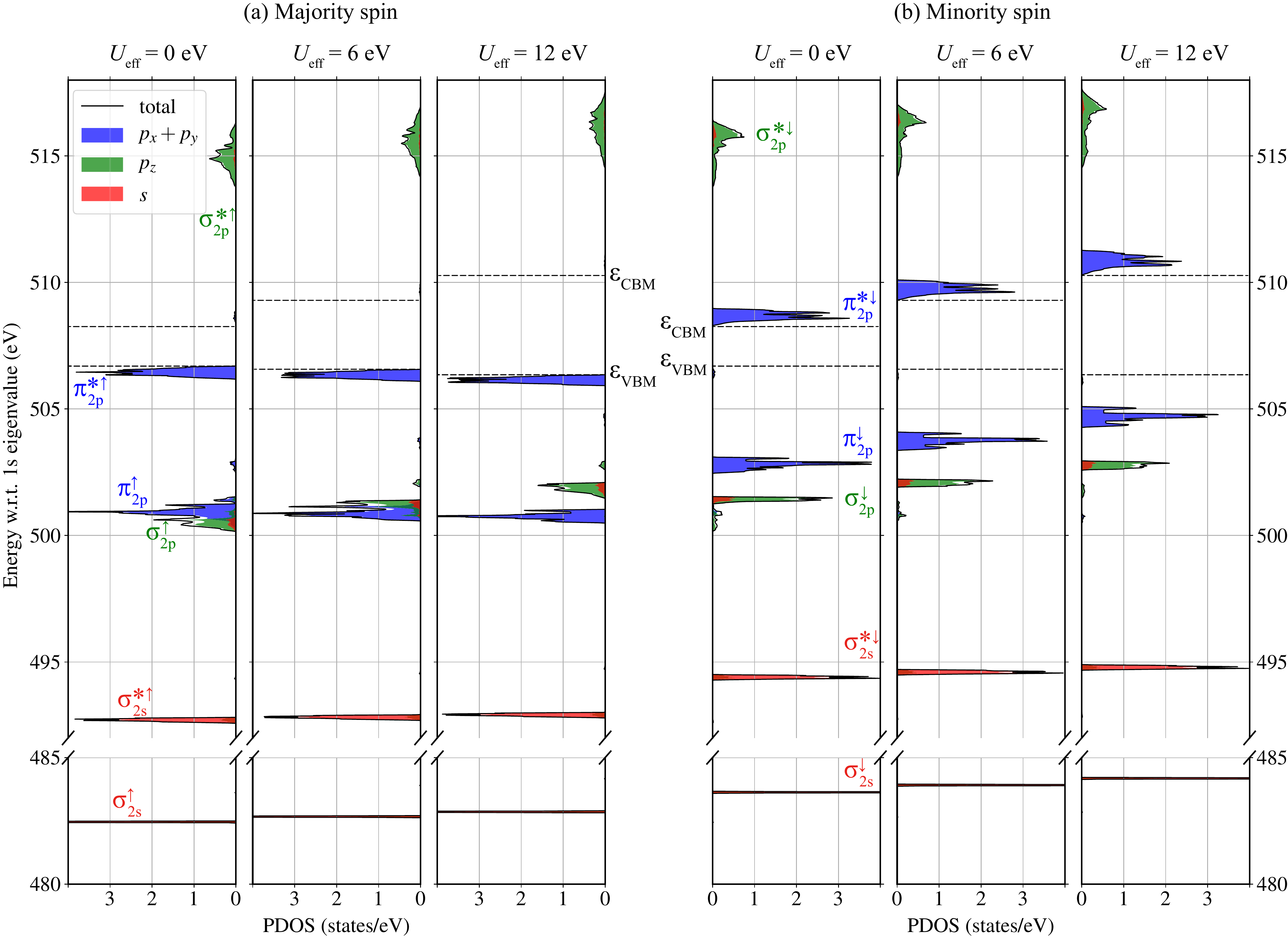}
\caption{\label{fig:PDOS} The majority-spin (a) and minority-spin(b) PDOS of an O$_2$ molecule in the $\alpha$-O$_2$ structure calculated using the vdW-DF-revPBE functional for various $U_\text{eff}$ values. 
The peaks are labeled according to
the symmetry of the corresponding O$_2$ molecular orbital. The valence band maximum ($\epsilon_\text{VBM}$) and conduction band minimum ($\epsilon_\text{CBM}$) are indicated by dashed horizontal lines.}

\includegraphics[width=1.5\columnwidth]{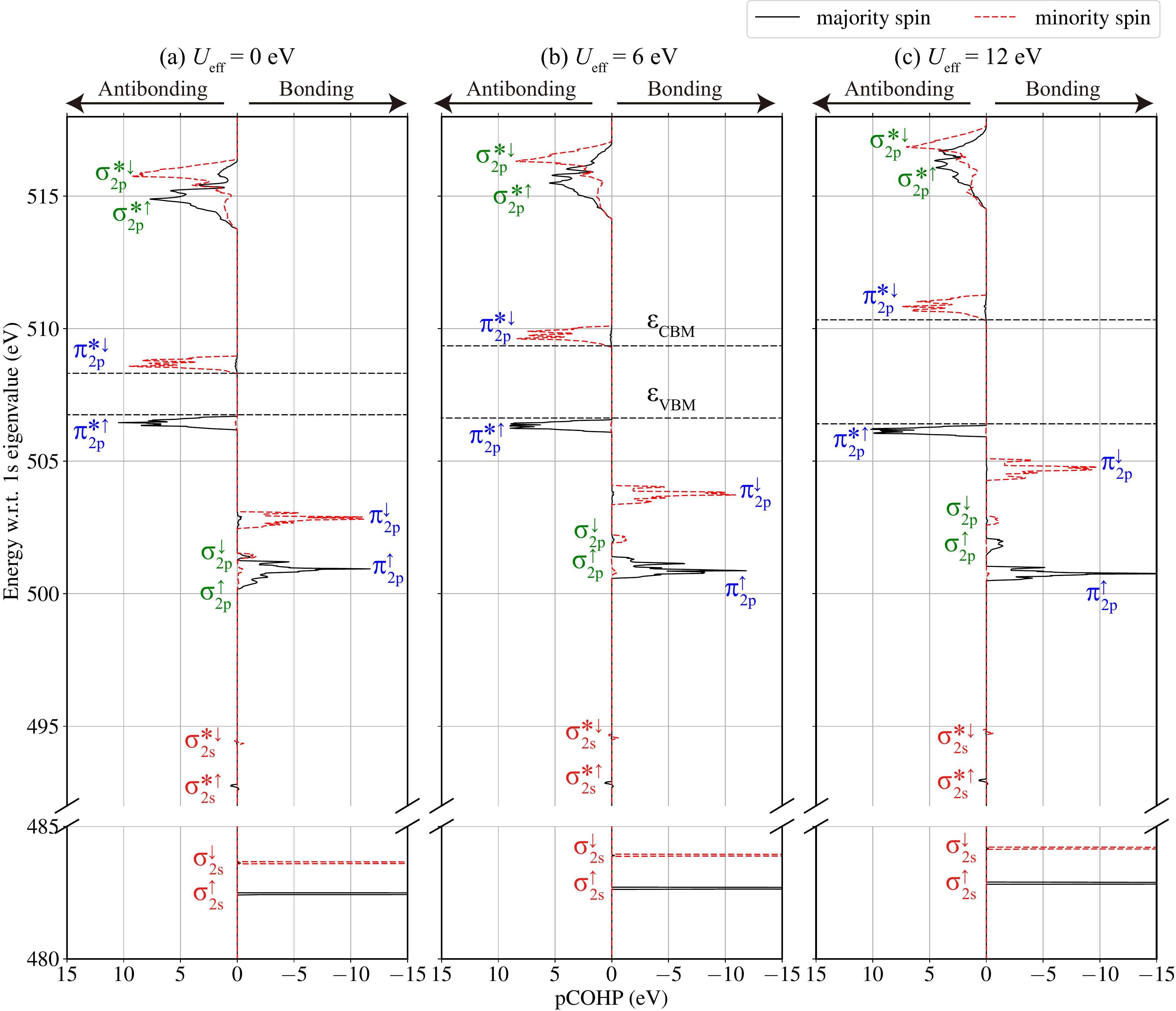}
\caption{\label{fig:pCOHP_intra} The pCOHP of the intramolecular O--O bond (see Fig.~\ref{fig:alphaO2}) for various $U_\text{eff}$ values. 
The peaks are labeled according to
the symmetry of the corresponding molecular orbital. The valence band maximum ($\epsilon_\text{VBM}$) and conduction band minimum ($\epsilon_\text{CBM}$) are indicated by dashed horizontal lines.}
\end{figure*}
%

\begin{figure*}
\centering
\includegraphics[width=1.5\columnwidth]{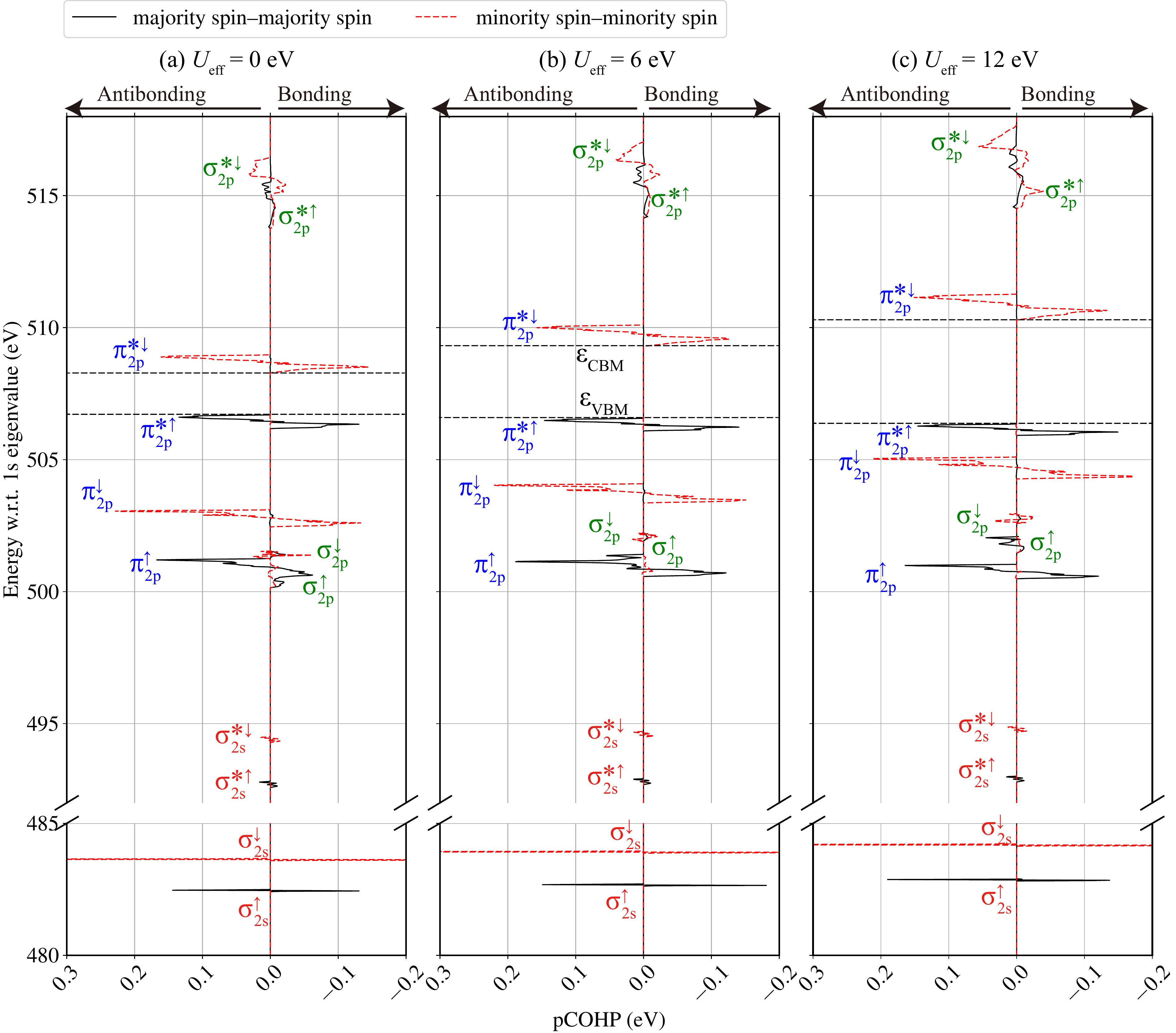}
\caption{\label{fig:pCOHP_FM} The pCOHP of the intermolecular FM O--O bond (see Fig.~\ref{fig:alphaO2}) calculated using the vdW-DF-revPBE functional for various $U_\text{eff}$ values. 
The peaks are labeled according to
the symmetry of the corresponding molecular orbital. The valence band maximum ($\epsilon_\text{VBM}$) and conduction band minimum ($\epsilon_\text{CBM}$) are indicated by dashed horizontal lines.}

\centering
\includegraphics[width=1.5\columnwidth]{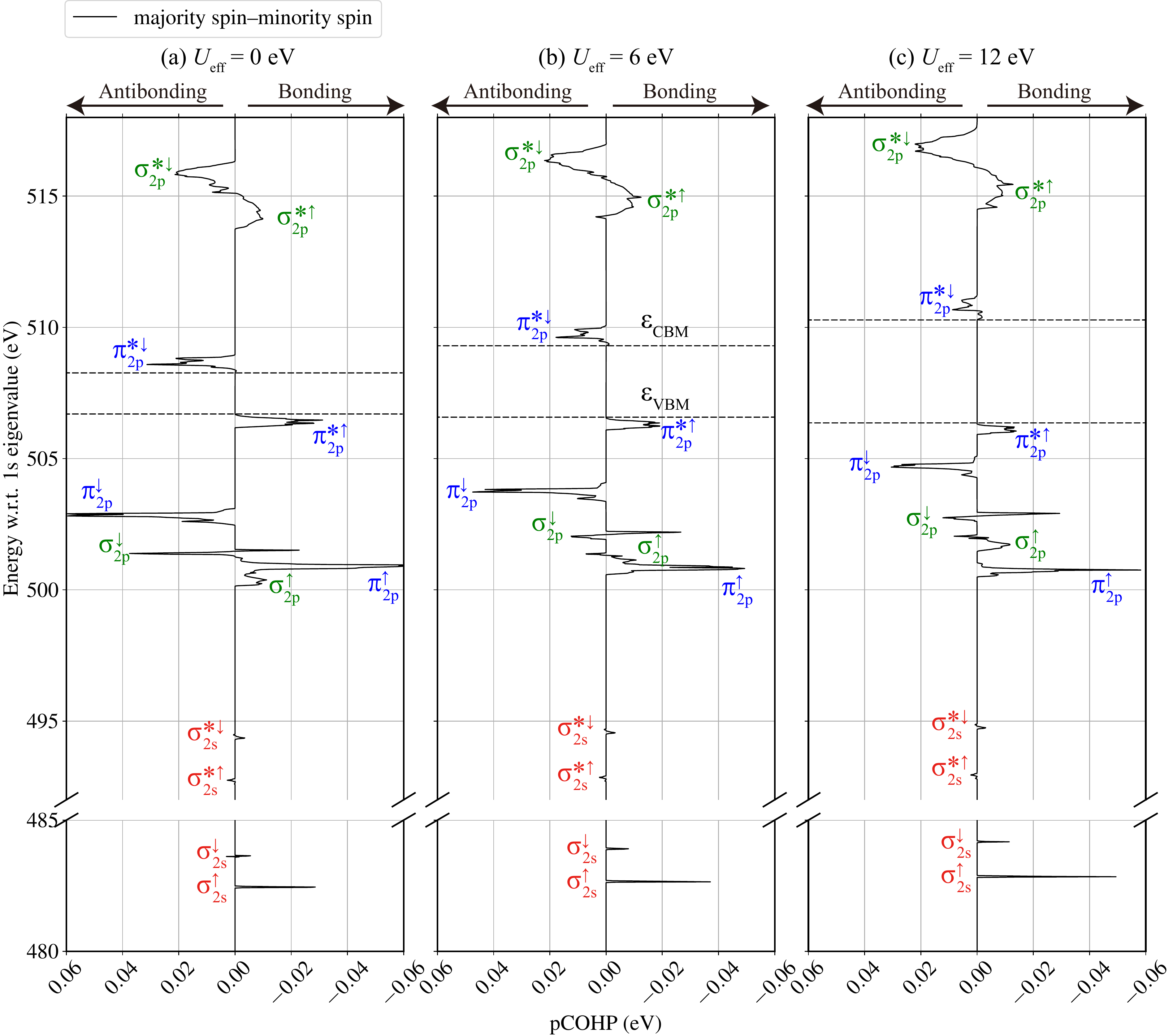}
\caption{\label{fig:pCOHP_AFM} The pCOHP of the intermolecular AFM O--O bond (see Fig.~\ref{fig:alphaO2}) calculated using the vdW-DF-revPBE functional for various $U_\text{eff}$ values. 
The peaks are labeled according to
the symmetry of the corresponding molecular orbital. The valence band maximum ($\epsilon_\text{VBM}$) and conduction band minimum ($\epsilon_\text{CBM}$) are indicated by dashed horizontal lines.}
\end{figure*}

\begin{figure*}
\centering
\includegraphics[width=2\columnwidth]{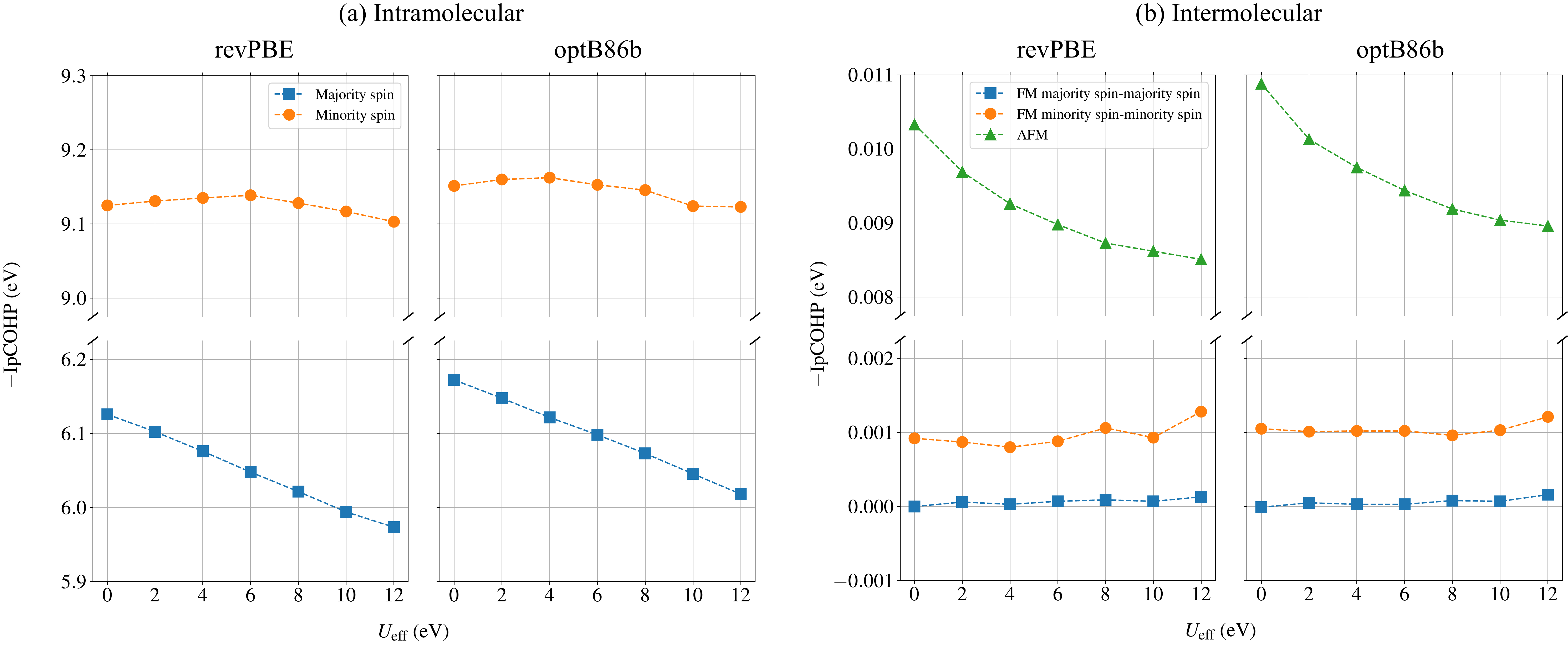}
\caption{\label{fig:IpCOHP} The $U_\text{eff}$-dependence of the $\text{IpCOHP}(\epsilon_\text{F})$ between intramolecular (a) and intermolecular (b)
 antiferromagnetic (AFM) and ferromagnetic (FM) intermolecular neighbor atoms calculated using vdW-DF-revPBE and vdW-DF-optB86b.}
\end{figure*}

\begin{figure*}
\centering
\includegraphics[width=2\columnwidth]{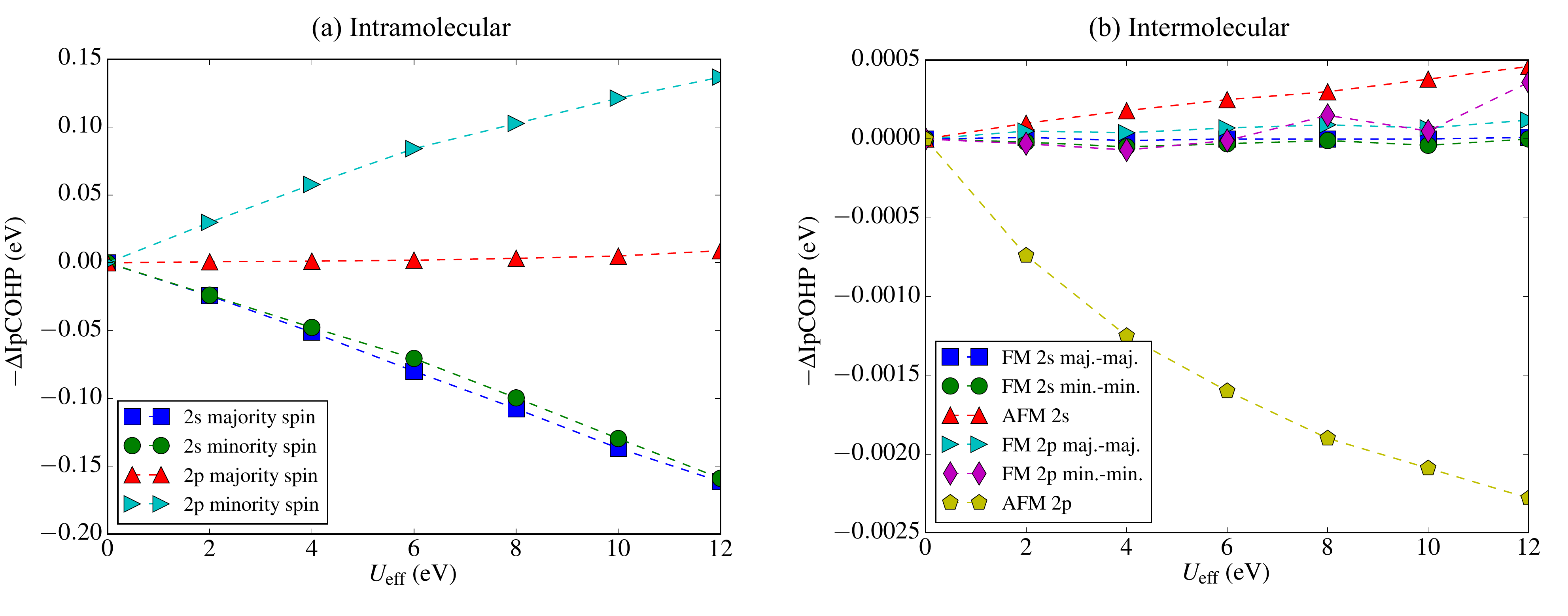}
\caption{\label{fig:dIpCOHP} The change in the IpCOHP of the intermolecular bond
integrated over the $2s$ (from 480 eV to 495 eV in Figs.~\ref{fig:pCOHP_intra},~\ref{fig:pCOHP_FM}, \ref{fig:pCOHP_AFM}) and $2p$ 
(from 495 eV up to $\epsilon_\text{VBM}$) manifolds for each spin channel as a function of  $U_\text{eff}$. Results are presented only for vdW-DF-revPBE.}
\end{figure*}

As shown above, tuning the single $U_\text{eff}$ parameter in vdW-DF$+U$
turns out to work surprisingly well in improving {\it all} lattice
parameters in the monoclinic $\alpha$ phase. In the following, we set
out to correlate this behavior with the effect of $U_\text{eff}$ on the
electronic structure and chemical bonding. 

The $U_\text{eff}$-dependence of the projected density of states (PDOS) on one of the oxygen atoms is shown in Fig.~\ref{fig:PDOS} for
vdW-DF-revPBE. The $U_\text{eff}$-dependence is also shown for the
pCOHP of the intramolecular O--O bond in Fig.~\ref{fig:pCOHP_intra} and for the pCOHP of ferromagnetic (FM) and antiferromagnetic (AFM) 
pairs across neighboring molecules in Figs.~\ref{fig:pCOHP_FM} and \ref{fig:pCOHP_AFM}, respectively. The optB86b results (not shown) look very similar
except for an upward shift in energy of about 0.5 eV measured from the 1s core level. 
In the usual
molecular orbital theory for the oxygen molecule, the $2s$ orbitals of each atom interact with each other to form bonding $\sigma_{2s}$
and antibonding $\sigma^\ast_{2s}$ molecular orbitals, while the $p_z$ orbitals form bonding $\sigma_{2p}$ and antibonding $\sigma_{2p}^\ast$ and $p_x$ and $p_y$
orbitals form bonding $\pi_{2p}$ and antibonding $\pi_{2p}^\ast$ orbitals (note that we have taken the $z$ axis to be parallel to the intramolecular O--O bond). The highest occupied molecular orbitals
are the two degenerate $\pi_{2p}^\ast$ orbitals which are each singly occupied in the triplet ground state leading to molecular magnetism.
The PDOS results of the vdW-DF$+U$ calculations (Fig.~\ref{fig:PDOS}) are basically consistent with this picture. That is, the nearest-neighbor O--O pairs 
in $\alpha$--O$_2$ retain most of their molecular character, although there is some broadening of each peak due to intermolecular interaction.

Figure \ref{fig:PDOS} shows that the $\sigma_{2p}$ states for
both spin channels and the $\pi_{2p}$/$\pi^\ast_{2p}$ states for the minority spin channel
shift up in energy with increasing $U_\text{eff}$, while the
energies of the majority-spin $\pi_{2p}$/$\pi^\ast_{2p}$ orbitals 
move down very slightly with respect to the core 1s level. 
As a result, the energy gap $\Delta E$ between the majority-spin $\pi_{2p}^\ast$ states comprising the valence band maximum (VBM)
and the minority-spin $\pi_{2p}^\ast$ states comprising the conduction band minimum (CBM) increases with
$U_\text{eff}$ as expected. 

\begin{figure*}
\centering
\includegraphics[width=1.75\columnwidth]{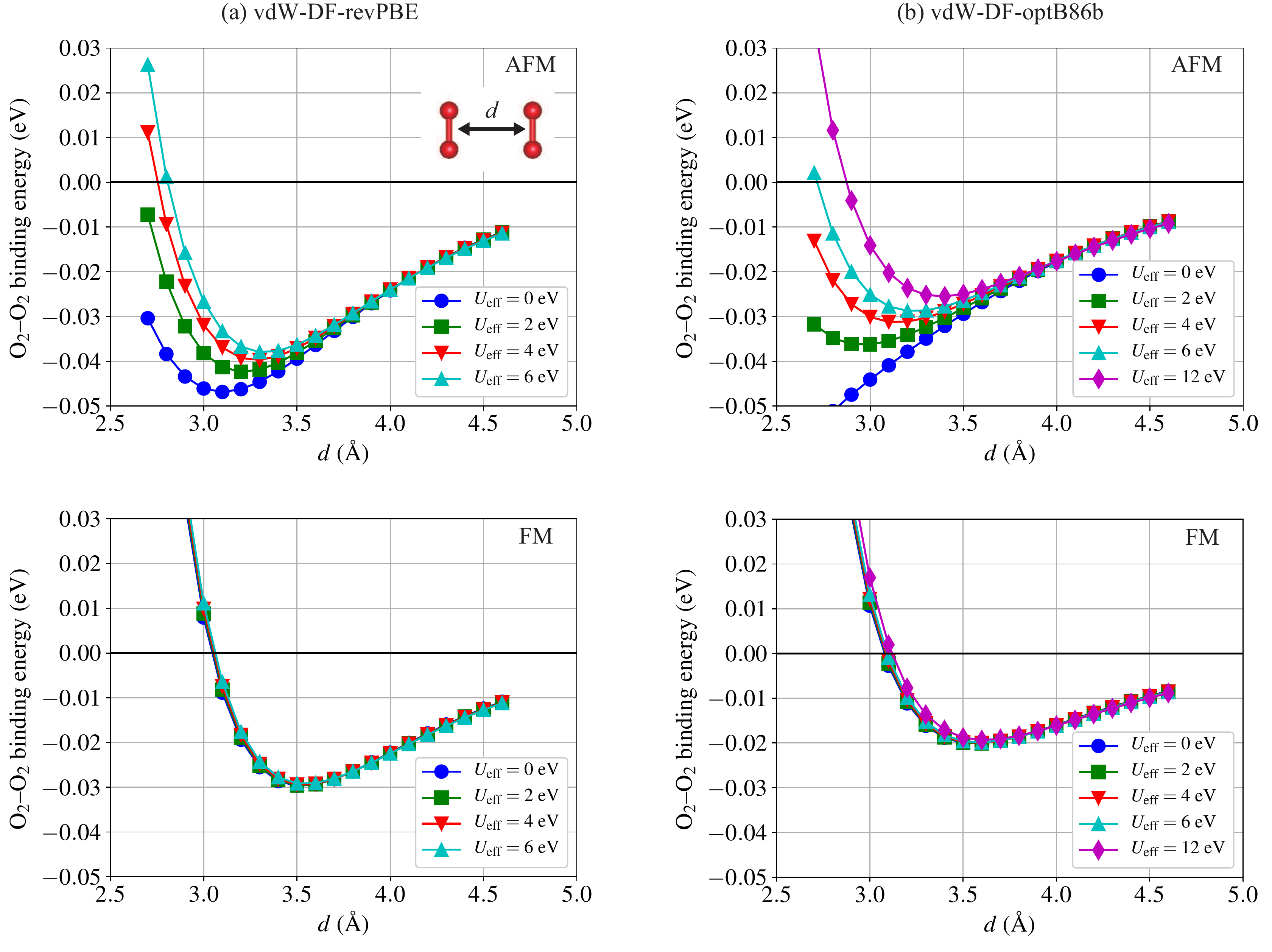}
\caption{\label{fig:O2O2bind} The O$_2$--O$_2$ binding energy as a function of distance $d$ for antiferromagnetic (top) and ferromagnetic (bottom) molecule pairs calculated using vdW-DF-revPBE$+U$ (a) and vdW-DF-optB86b$+U$. The intramolecular O--O distance is fixed at 1.25 \r{A}.}
\end{figure*}

The pCOHP of the intramolecular bond is shown in Fig.~\ref{fig:pCOHP_intra}.  In the vdW-DF calculations examined here, the $\sigma^\ast_{2s}$ and $\sigma_{2p}$ 
orbitals are nearly nonbonding, presumably due to hybridization with each other (see PDOS in Fig.~\ref{fig:PDOS}). The total bonding strength measured by the negative of the IpCOHP (Fig.~\ref{fig:IpCOHP} left)
weakens with increasing $U_\text{eff}$,  resulting in the increase in the intramolecular bond length (Fig.~\ref{fig:U-lat}). This is the expected behavior, as the Hubbard $U$
encourages localization of electrons and discourages bonding. However, when we decompose the IpCOHP into $2s$ and $2p$ manifolds (Fig.~\ref{fig:dIpCOHP} left), we find that the $U_\text{eff}$, 
which is added only on the $2p$ onsite term, results in the weakening of the $\sigma_{2s}$ bond while having a much smaller effect on the $2p$ bonds. We expect this to be due
to the electrostatic screening between 2p and 2s electrons that is perturbed by the addition of the $2p$ on-site $U_\text{eff}$, although we do not have a concrete explanation at this moment
for the resulting behavior.

Next, we examine the intermolecular bonding. When comparing AFM and FM intermolecular pairs, the pCOHP and the IpCOHP 
show distinctly different
features (see Fig.~\ref{fig:pCOHP_FM}-\ref{fig:pCOHP_AFM} and \ref{fig:IpCOHP} right). The pCOHP between the AFM pairs shows peaks near the Fermi level that are smaller in magnitude
than that between FM pairs, indicating a more moderate interaction between AFM pairs. However, the negative of the integrated IpCOHP
(Fig.~\ref{fig:IpCOHP}), which is a measure of the total bond strength, is
much larger for AFM pairs. This is because the Fermi
level is located between the bonding and antibonding orbitals formed by hybridization between $\pi^\ast$ orbitals of
the AFM pairs (see Fig.~\ref{fig:pCOHP_AFM}), while no such feature exists for FM pairs 
(i.e., all bonding-antibonding orbital pairs are below the Fermi level).  This in turn originates from the fact that in AFM
pairs, the majority spin orbitals
of one of the O$_2$ pairs interacts with the minority spin orbitals of the other O$_2$
molecule, while in FM pairs, the majority spin orbitals of one molecule
interact with majority spin orbitals of the other molecule with the same
energy, and vice versa. Thus, in the calculated IpCOHP, 
a very weak bonding character is observed for FM pairs regardless of the $U_\text{eff}$ value, while
the bonding character is more significant for AFM pairs and the decrease
due to $U_\text{eff}$ is also much more prominent. Such small bonding character between FM pairs is a manifestation of the Pauli exclusion
principle; fully occupied orbitals do not form bonds with each other. We also note that the decrease in the bonding strength of AFM pairs originates mainly from the weakening 
of the bonding between $2p$ orbitals of neighboring molecules; the $2s$ orbitals, on the other hand, act to slightly strengthen the bond with increasing $U_\text{eff}$.
The decomposition of the intermolecular 
IpCOHP into $2s$ and $2p$ manifolds (Fig.~\ref{fig:dIpCOHP} right) shows that unlike the intramolecular case (Fig.~\ref{fig:dIpCOHP} left), most of the change
in the bonding strength comes from the $2p$ manifold, although the $2s$ manifold also shows non-negligible change vs. the $U_\text{eff}$ value.

We also note that when comparing the IpCOHP for revPBE and optB86b at each $U_\text{eff}$ value, the latter shows stronger bonding between AFM pairs (Fig.~\ref{fig:IpCOHP}); this is in line with the usual trend of
vdW-DF-optB86b to predict smaller bond lengths compared to vdW-DF-revPBE \cite{Berland2015}. Thus, vdW-DF-optB86b requires a larger $U_\text{eff}$ value for weakening the AFM bonding to match
the experimental structure. It is also worth noting that the IpCOHP at the optimal $U_\text{eff}$ values for reproducing the experimental structure
($U_\text{eff} = 5$ eV for revPBE and 12 eV for optB86b) are very similar; this indicates that the bond strength evaluated  using IpCOHP correlates closely with the resulting structure regardless of the functional
approximation.

For further confirmation that the vdW-DF$+U$ approach is indeed effective in correcting the overstabilization of the bonding between antiferromagnetic pairs, we examine the binding energy of two parallel O$_2$ molecules in vacuum calculated as
$E_{\text{O}_2\text{--}\text{O}_2} -2 E_{\text{O}_2}$, where $E_{\text{O}_2\text{--}\text{O}_2}$ is the energy of the bonded O$_2$ pair and $E_{\text{O}_2}$ is the energy of an isolated O$_2$ molecule (Fig.~\ref{fig:O2O2bind}).
It is clearly seen that increasing the $U_\text{eff}$ parameter results in increased bonding distance and decreasing bonding energy for AFM pairs, while it has virtually no effect on FM pairs.
Thus, the $+U$ approach provides the desired correction for the originally overestimated magnetic interaction $J$.

At this point, we may reconcile why vdW-DF-optB86b results in
lattice constants in better agreement with experiment compared
to vdW-DF-revPBE when the $U_\text{eff}$ parameter is optimized.
The key point is that the interaction between FM
pairs need to be described correctly, since it cannot be tuned
by the $U_\text{eff}$ parameter as seen in
Fig.~\ref{fig:O2O2bind}.
The vdW-DF-optB86b result [Fig.~\ref{fig:O2O2bind} (b)] for
the FM O$_2$--O$_2$ dimer is closer to
highest-accuracy quantum chemistry calculations available in the
literature for the quintet state of the dimer, which predicts an
interaction energy of $\sim 14$ meV \cite{Novillo2010}. The
better result may be due to the fact that the optB86b exchange
is similar to the exchange employed in the newer vdW-DF-{\it cx}
\cite{Berland2014}, where the exchange functional is constructed
to be more consistent with the
underlying justification of vdW-DF based on adiabatic connection
\cite{Berland2015}. On the other hand, the reason for the worse
performance at $U_\text{eff} = 0$ for the AFM
pairs is difficult to track down. The vdW-DF functional
form may simply be inept at describing such a system, and the
seemingly better performance of the vdW-DF-revPBE may not be for
the correct reasons. In fact, the strongly correlated and
multireference nature of the singlet (i.e., AFM) state of the
O$_2$ dimer is a challenge even for
multireference quantum chemistry approaches
\cite{Bartolomei2010}.

Summarizing the above, we have achieved acceptable levels of accuracy in describing the $\alpha$ phase
so far unattained in the literature employing DFT-based methods. 
This means that vdW-DF$+U$ is likely to be a viable approach for semiquantitative examination of
the oxygen temperature-pressure-magnetic field phase diagram. In the following, we apply this method to
examination of the newly-discovered ferromagnetic $\theta$ phase which appears at high magnetic fields.
A more systematic study of the phase diagram is deferred to future works.

%
%

\subsection{Examination of the candidate Pa$\bar{3}$ structure for the $\theta$ phase}
\begin{figure}[tb]
\centering
\includegraphics[width=0.8\columnwidth]{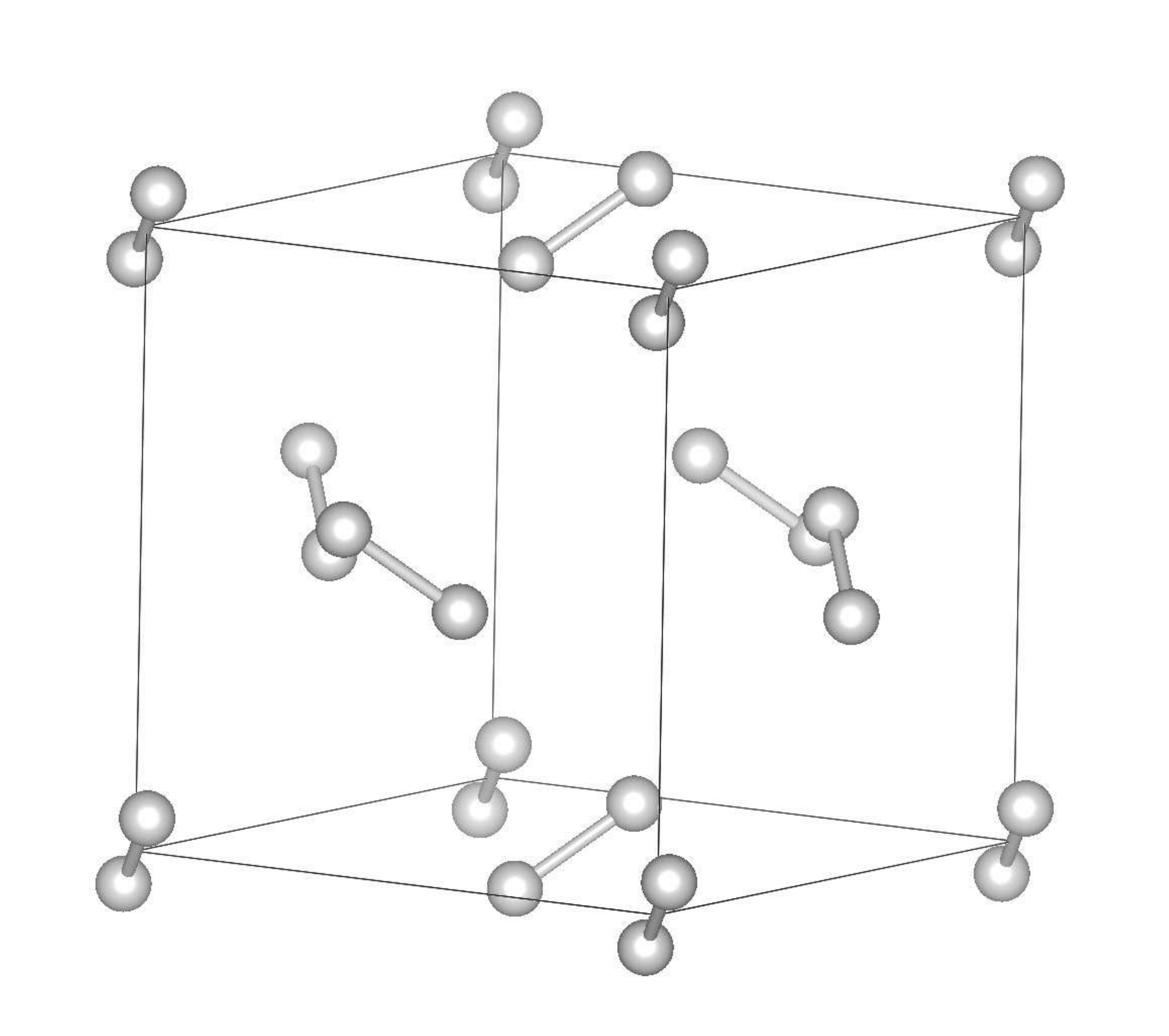}
\caption{\label{fig:thetaO2} The proposed Pa$\bar{3}$ structure for the magnetic field-induced $\theta$ phase of solid 
O$_2$.}
\end{figure}
The 
high-magnetic field experiments by Nomura {\it et al.} \cite{Nomura2014,Nomura2015} show an increase in the 
magnetization per O$_2$ molecule to over 1.5$\mu_\text{B}$, suggesting that the newly discovered phase is 
ferrimagnetic or ferromagnetic.  It is known that ferromagnetic O$_2$-O$_2$ dimers are unstable in the parallel-aligned 
geometry seen in the $\alpha$ phase examined above; to minimize Pauli repulsion between 
electrons with the same spin, ferromagnetic O$_2$-O$_2$ dimers instead tend towards canted or crossed  
arrangements \cite{Hemert1983,Bussery1993,Nozawa2002}. Moreover, magnetotransmission experiments exhibit 
high transmission intensity in the $\theta$ phase compared to $\alpha$ and $\beta$ phases. This means that scattering of incident light 
at domain boundaries is decreased in the $\theta$ phase, and it is suggested that this is due to decrease of
crystalline anisotropy, i.e., formation of a cubic phase. Based on these observations, Nomura et al. suggest that the 
structure of $\theta$ phase is the cubic Pa$\bar{3}$ stucture shown in Fig.~\ref{fig:thetaO2}, which is also the structure of 
low-temperature phases of  CO$_2$, N$_2$, and N$_2$O \cite{Manzhelii1997}. 

\begin{figure}[tb]
\centering
\includegraphics[width=\columnwidth]{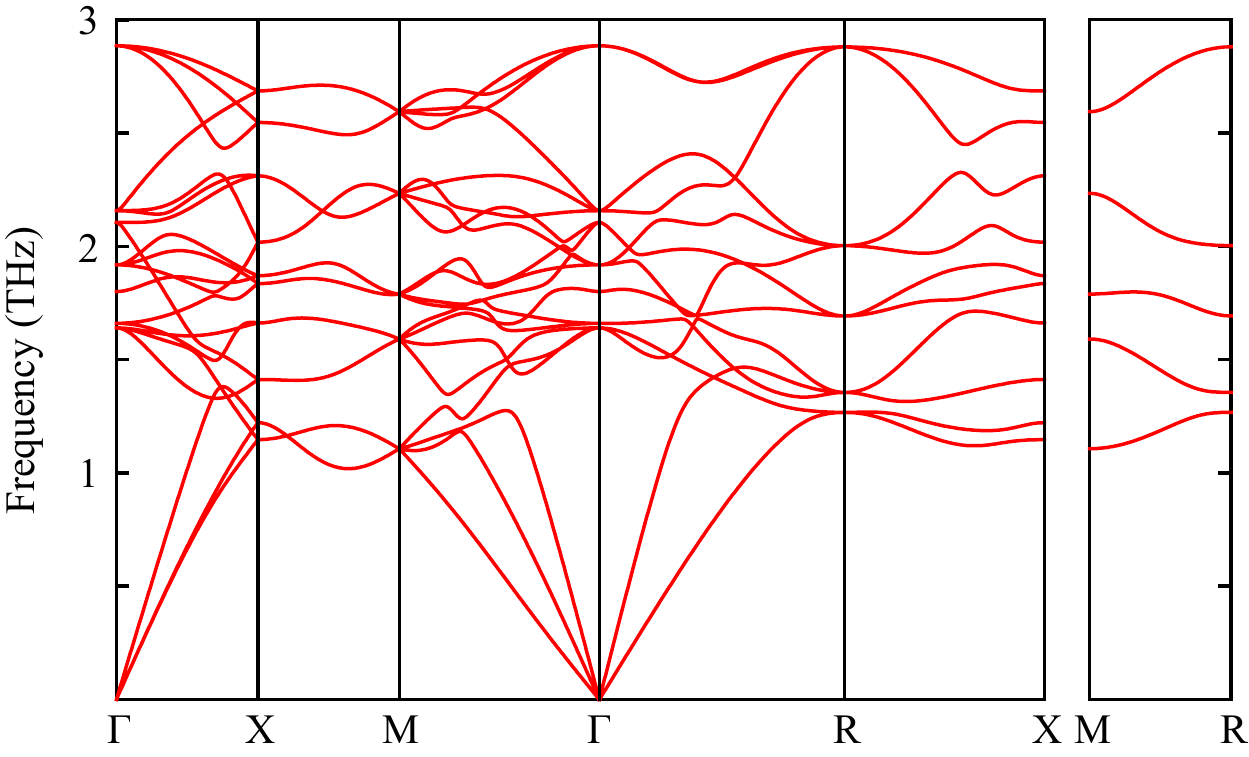}
\caption{\label{fig:phonon} The phonon dispersion of the Pa$\bar{3}$ structure of solid oxygen calculated using vdW-DF-revPBE$+U$. There are additional phonon modes around 40 THz corresponding to the O$_2$ stretching mode (not shown).}
\end{figure}

\begin{table*}[tb]
  \centering
  \caption{The lattice parameter $a$, internal parameter $x$ of the 8c position, and the energy difference $\Delta E_{\alpha \rightarrow \theta} = E_\theta - E_\alpha$ per O$_2$  molecule in the Pa$\bar{3}$ structure calculated using various functionals.}
\begin{ruledtabular}
    \begin{tabular}{lccc}
            & $a$ (\r{A}) & $x$ & $\Delta E_{\alpha \rightarrow \theta}$ (meV/O$_2$) \\
\hline
    vdW-DF-revPBE  & 5.36    & 0.066    & 37 \\
    vdW-DF-revPBE$+U$ ($U_\text{eff}=5$ eV) & 5.39    & 0.067    & 13 \\
    vdW-DF-optB86b$+U$ ($U_\text{eff}=12$ eV) & 5.25    & 0.070    & 8  \\

    \end{tabular}%
\end{ruledtabular}
  \label{Tab:theta}%
\end{table*}%

To confirm that this structure is at least locally stable, we perform structural relaxation starting from the Pa$\bar{3}$ 
structure and a ferromagnetic electron configuration. We compare the vdW-DF-revPBE$+U$ functional with $U_\text{eff}$ of 5 eV and 0 eV, as well
as the vdW-DF-optB86b$+U$ functional with $U_\text{eff} = 12$ eV.

The resulting structure is found to be quite similar in all three cases (Table \ref{Tab:theta}), although the vdW-DF-optB86b$+U$ predicts a
smaller lattice constant compared to vdW-DF-revPBE in line with the usual trend mentioned in Sec.~IIIB.
We also calculated the phonon band structure of the fully relaxed Pa$\bar{3}$ structure (Fig.~\ref{fig:phonon}) and found that there are no imaginary modes, i.e., it 
was confirmed that this structure is stable in the ferromagnetic electron configuration. 

Comparing the $U_\text{eff}=0$ and 5 eV cases for vdW-DF-revPBE, we find that the $U$ parameter has virtually no effect on the
predicted structure parameters. 
This is because as noted above, the $U$ parameter has minimal effect on the ferromagnetic state whose 2$p$ electrons are already localized at  $U=0$ 
eV due to Pauli repulsion between electrons with same spin.

We may make a rough estimation that the free energy of the $\alpha$ phase depends little on the external magnetic field due to cancellation between magneto-expansion and the Zeeman energy term $-g\mu_\text{B} S B_\text{ext}$ as it becomes partially ferromagnetic. In this case, we can relate the total energy difference at zero field to 
the Zeeman energy at the phase transition point as
$E_\alpha - E_\theta = -
g\mu_\text{B} B_\text{crit}$, where $g \sim 2$ is the electron spin $g$-factor, $\mu_\text{B}$ is the Bohr magneton, 
and $B_\text{crit}$ is the critical field for the transition. From the calculated energies (Table \ref{Tab:theta}), vdW-DF-revPBE
predicts a critical field of $\sim 300$ T, vdW-DF-revPBE$+U$  with $U_\text{eff}=5$ eV predicts $\sim 110$ T, and vdW-DF-optB86b$+U$
with $U_\text{eff}=12$ eV predicts $\sim 70$ T.
The latter two are in decent agreement with  $B_\text{crit} \sim 100$ T found in experiment, suggesting (although the evidence is still rather circumstantial) that Pa$\bar{3}$
is indeed the structure of the $\theta$ phase discovered by Nomura et al. The higher $B_\text{crit}$ predicted by vdW-DF without Hubbard $U$ is most likely due to the relative overstabilization of the $\alpha$ phase originating from 
the overestimation of the antiferromagnetic interaction discussed in preceding sections. 

\section{Conclusion}
In this work, we showed that the addition of the Hubbard $U$ energy term to vdW-DF functionals gives the best description of the structure of $\alpha$-O$_2$ obtained thus far in the literature using DFT-based methods. All lattice parameters ($a$, $b$, $c$, and $\beta$) and the intramolecular bond length $l_{\text{O}_2}$ in the monoclinic phase improve with addition of the $+U$ term, and this is attributed to the correction of the overbinding of O$_2$ pairs with antiparallel spins compared to parallel spins. 
We also applied this approach to the proposed Pa$\bar{3}$ structure of the high magnetic field $\theta$ phase, and confirmed that the energetics seem to be in line with experiment and that the structure is stable.
In a broader context, we reiterate the notion first given in Ref.~\onlinecite{Obata2015a}: this approach is a clear step forward in quantitative prediction of magnetic and structural properties in systems where vdW and spin-spin interaction compete, such as molecular magnets and metalorganic systems. Application of this approach to study of molecular spintronics is highly anticipated.


%

\begin{acknowledgments}
The authors would like to thank Yasuhiro Matsuda and Toshihiro Nomura of the Institute for Solid State Physics (ISSP), the University of Tokyo for fruitful discussion. The calculations were performed using the ISSP supercomputer system. Atomic structure figures were created using the visualization software VESTA \cite{Momma2008}.
\end{acknowledgments}

\bibliographystyle{apsrev4-1}
\bibliography{article}

\end{document}